          [2004/06/22 1.2 (KOF); 2001/04/25 1.1 (PWD)]

\documentclass[a4paper,twocolumn]{Gaia2004} 
\usepackage{times}      
\usepackage{epsfig}     
\usepackage{graphicx}     
\usepackage{natbib}     
\title{A Keck/HIRES Doppler Search for Planets Orbiting Metal-Poor Dwarfs}

\author[1,2]{A. Sozzetti}
\affil[1]{Department of Physics \& Astronomy, University of 
Pittsburgh, Pittsburgh, PA 15260 USA}
\affil[2]{Harvard-Smithsonian Center for Astrophysics, 60 Garden Street, 
Cambridge, MA 02138 USA}
\author[2]{D. W. Latham}
\author[2]{G. Torres}
\author[2]{R. P. Stefanik}
\author[3]{A. P. Boss}
\author[4]{B. W. Carney}
\author[5]{J. B. Laird}
\affil[3]{Carnegie Institution of Washington, 
5241 Broad Branch Road, NW, Washington, DC 20015 USA}
\affil[4]{Department of Physics \& Astronomy,
University of North Carolina at Chapel Hill, Chapel Hill, NC 27599 USA}
\affil[5]{Department of Physics \& Astronomy,
Bowling Green State University, Bowling Green, OH 43403 USA}
\bibpunct{(}{)}{;}{a}{}{,}  

\begin{document}

\maketitle

\keywords{planetary systems; radial velocity; astrometry; 
stars: statistics}

\begin{abstract}
  
We present results from our ongoing spectroscopic search for 
giant planets within 1 AU around a well-defined sample of metal-poor 
stars with HIRES on the Keck 1 telescope. We have achieved an rms 
radial velocity precision of $\sim 8$ m/s over a time-span of 1.5 years. 
The data collected so far 
build toward evidence of the absence of very short-period ($< 1$ month) 
giant planets. However, about 7\% of the stars in our sample exhibits 
velocity trends indicative of the existence of companions. 
We place preliminary upper limits on the detectable companion mass 
as a function of orbital period, and compare them with the performance 
of ESA's future space-borne high-precision astrometric observatory Gaia. 

\end{abstract}

\section{Introduction}

With a present-day catalogue of well over 130 extrasolar planets
\footnote{See for example http://www.obspm.fr/encycl/encycl.html}, 
several important statistical properties of the sample are beginning 
to emerge. One of the most intriguing features unveiled so far, however, 
concerns the parent stars rather than the planets themselves. 
In particular, both the probability of a star to harbor a planet and 
some orbital properties of the latter appear to depend on the metallicity 
of the former. 

The metallicity distribution of stars with planets peaks at 
[Fe/H] $\simeq 0.3$, showing evidence of moderate metal-enrichment 
with respect to the average metallicity ([Fe/H] $\simeq -0.1$) of 
field dwarfs in the solar neighbourhood (Santos et al. 2001; 
Fischer et al. 2003; Santos et al. 2004). The evidence for higher planetary 
frequency around metal-rich stars has been 
confirmed based on observationally unbiased stellar samples. 
This trend seems to agree with the 
predictions from theoretical models of gas giant planet formation by 
core accretion (e.g., Ida \& Lin 2004). However, the 
alternative scenario of giant planet formation by disk instability 
(Boss 2002) is insensitive to the primordial metallicity of the 
protoplanetary disk, and, although not statistically significant, 
the possible evidence for bi-modality of the planet 
frequency distribution as a function 
of metallicity (Santos et al. 2004) suggests the existence 
of two different mechanisms for forming gas giant planets. 

Furthermore, despite potential biases introduced by the small-number 
statistics, the orbital periods of extrasolar planets seem to correlate 
with the metallicity of their parent stars (Sozzetti 2004, and references 
therein). In particular, close-in planets, on few-day orbits, are 
more likely to be found around metal-rich stars. If true, the correlation 
could reflect a dependence of migration rates on the amount of metals 
present in the disk (Livio \& Pringle 2003). Alternatively, it might be 
indicative of longer timescales for giant planet formation around 
metal-poor stars, and thus reduced chances for the protoplanets to undergo 
significant migration before the disk evaporates (Ida \& Lin 2004). 

Such questions can be addressed by comparing the frequency of gas giant 
planets and their properties between metal-rich and metal-poor stars.  
However, the low-metallicity stellar sample is at present too small to 
test but the most outstanding differences between such hypothetical 
populations. It is then crucial to provide a statistically significant, 
unbiased sample of metal-poor stars screened for giant planets. 
This can be achieved by means of both Doppler (Sozzetti et al. 2003a) 
and astrometric (Sozzetti et al. 2001, 2002, 2003b) surveys. 

\section{Selection Criteria of the Sample}

\begin{figure*}[!t]
\centering
$\begin{array}{cc}
\includegraphics[width=0.425\textwidth]{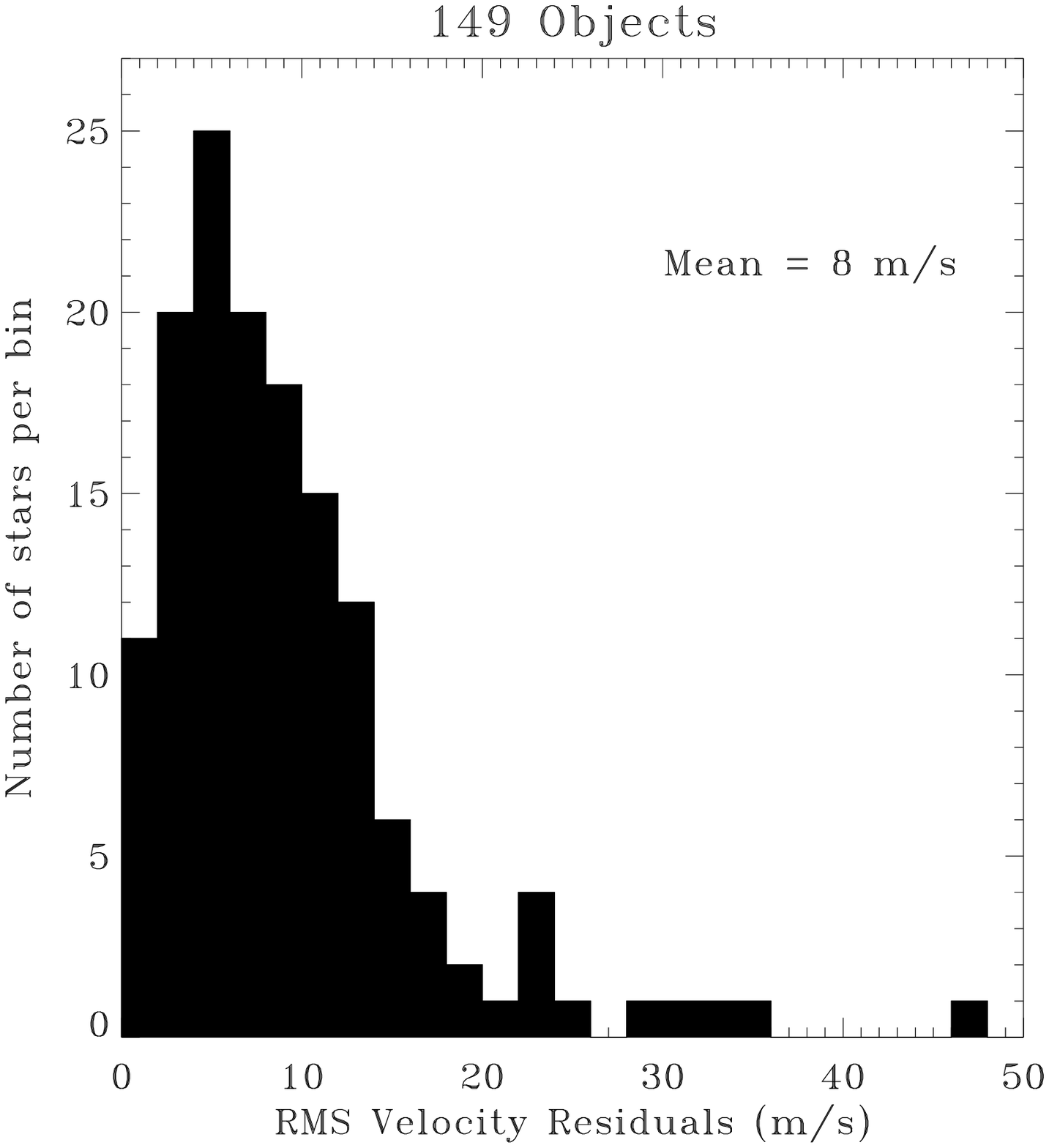} & 
\includegraphics[width=0.40\textwidth]{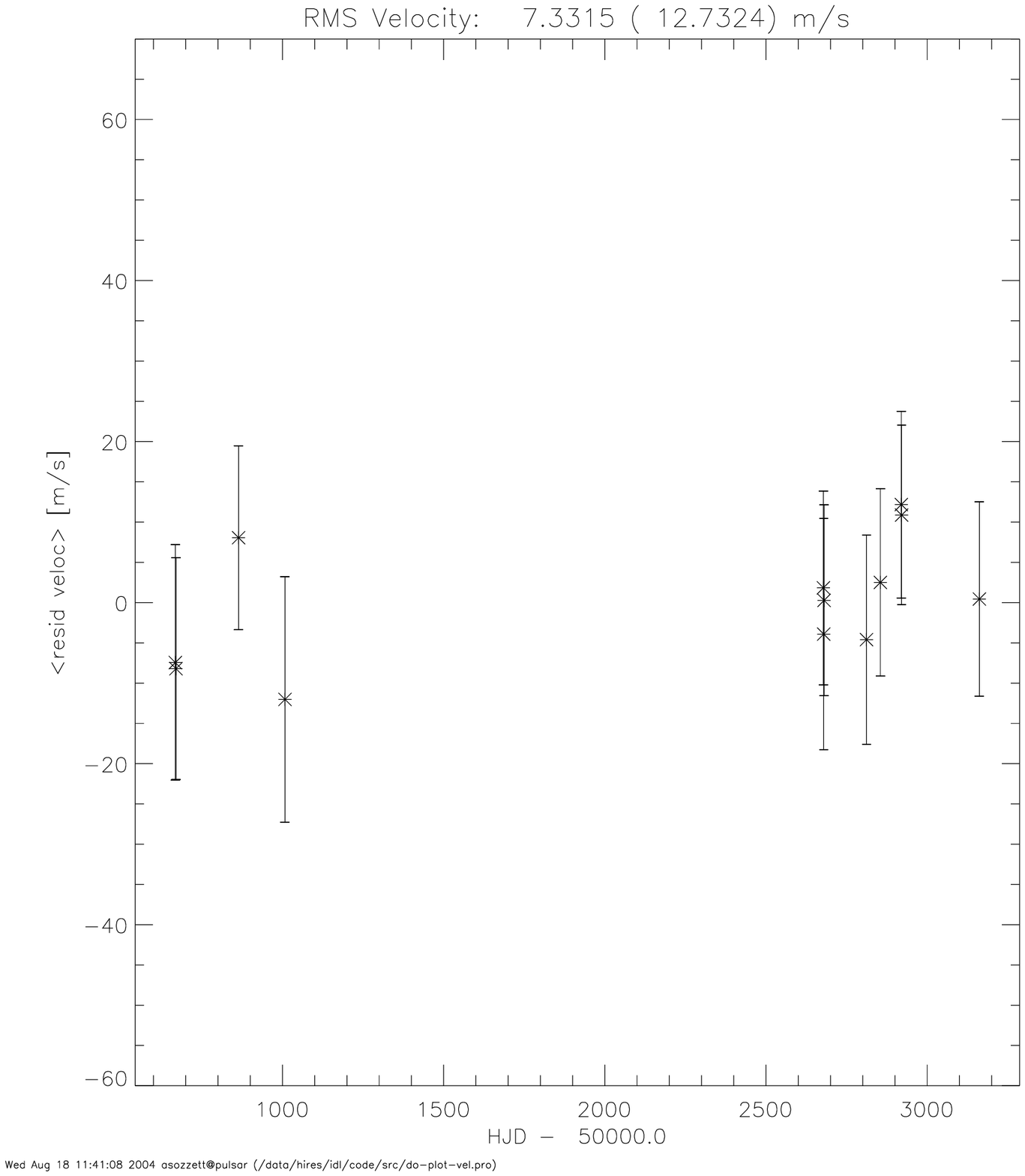} 
\end{array} $
\label{rmsdistr}
\caption{Left: rms velocity distribution for the full sample. A number 
of objects exhibiting significant radial velocity variations 
($> 50$ m/s) is not 
shown. Right: radial velocity as a function of time for a quiet 
star in our sample.}
\end{figure*}
In this project we are using the HIRES spectrograph on the Keck 1 telescope 
(Vogt et al. 1994) to search for planetary companions within 1 AU 
orbiting a sample of 200 metal-poor dwarfs. The sample has been drawn from 
the Carney-Latham and Ryan samples of metal-poor, high-velocity field stars 
(e.g., Carney et al. 1994; Ryan 1989). The stars have been selected not 
to have close orbiting companions in the stellar mass regime that might 
hamper the formation or survival of planets (Carney et al. 2001; 
Latham et al. 2002).  

Old stars have the advantage that they rotate slowly and have low
levels of chromospheric activity.  All of the stars in our sample
exhibit rotational velocities $V_\mathrm{rot}\leq 10$ km/s. 
Thus, velocity jitter due to astrophysical phenomena is not expected to be a
problem for this sample. 
However, metal-poor stars have weak absorption lines in comparison to 
their solar-metallicity counterparts. The lines also grow weaker as the
effective temperature rises. Furthermore, very metal-poor stars are
rare, and therefore they tend to be distant and faint. In order to 
characterize the behavior of the radial velocity precision as a 
function of stellar metallicity [Fe/H], effective temperature $T_\mathrm{eff}$, 
and visual magnitude $V$ (assuming non-rotating, inactive stars), we have
run simulations utilizing the CfA library of synthetic stellar spectra 
(Sozzetti et al. 2003a). In light of those results, we have refined 
our sample of 200 metal-poor dwarfs from the Carney-Latham and Ryan
surveys by selecting objects in the metallicity range 
$-2.0 \leq$ [Fe/H] $\leq -0.6$, and utilized the following magnitude and 
temperature cut-offs: $V\leq 11.5$ and $T_\mathrm{eff}\leq 6250$ K. 

Based on our experience with 
solar neighborhood G dwarfs observed with HIRES for the G Dwarf Planet
Search Program (Latham 2000), we have set an initial threshold of 20 m/s
precision for planet detection, and have computed the relative
exposure times needed to achieve such precision, for each star in our
sample. 


Finally, our sample-size is large enough that a null result, 
i.e. no detections, would be significant. The frequency of giant
planets within 1 AU around F-G-K dwarfs is 
$f \simeq 3$-4\% (e.g., Santos et al. 2004). In
order for the failure to detect any planetary companions to be
significant at the 3-$\sigma$ level (corresponding to a probability of
0.0027), we need to survey a sample of $N$ stars, where
$(1-f)^N = 0.0027$, which is satisfied for $N = 194-145$. Our
sample of 200 metal-poor stars should eventually provide a robust
3-$\sigma$ null result in case of no detections.

\section{Results}


\begin{figure*}[!t]
\centering
$\begin{array}{ccc}
\includegraphics[width=0.30\textwidth]{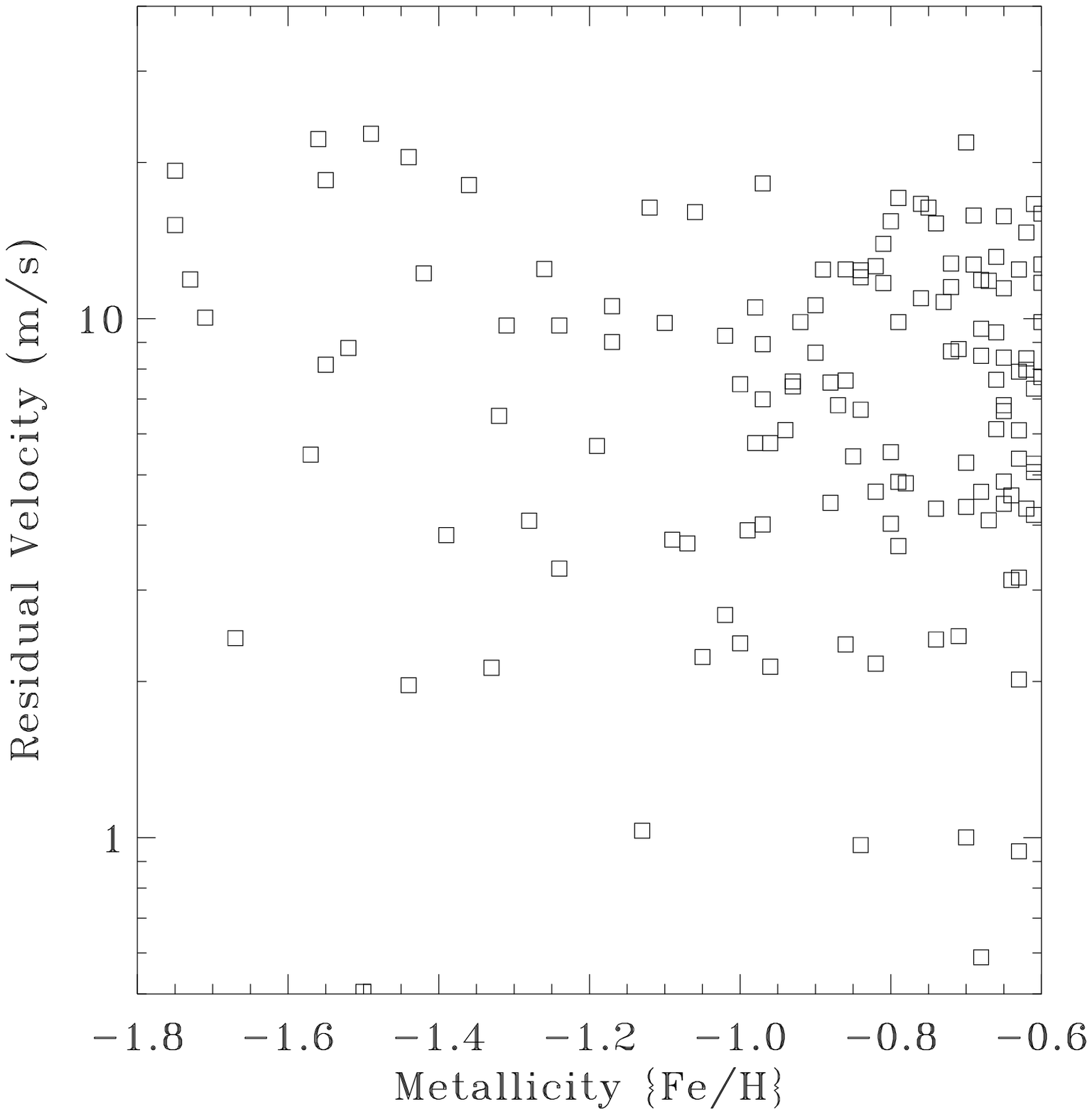} & 
\includegraphics[width=0.30\textwidth]{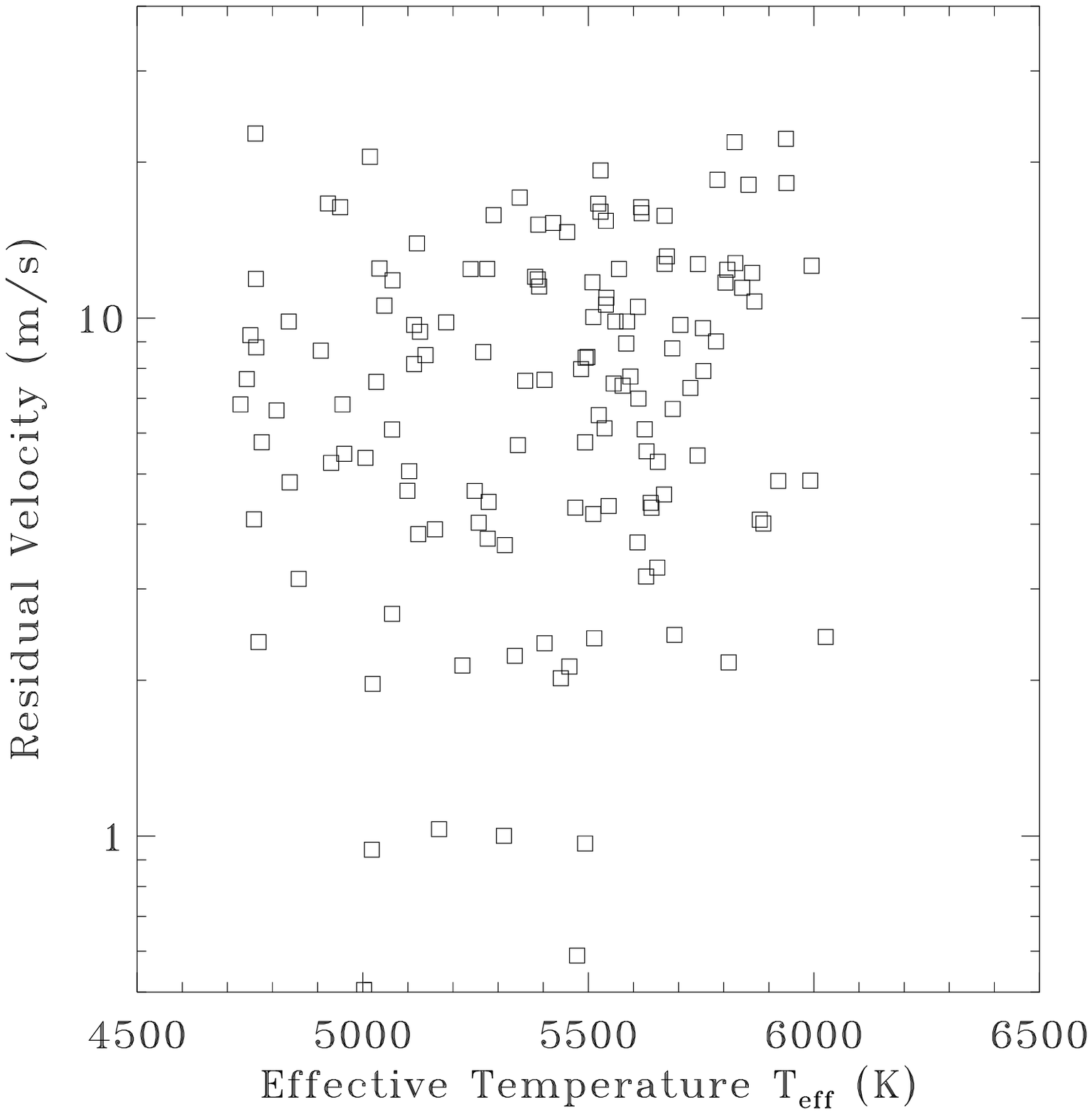} & 
\includegraphics[width=0.30\textwidth]{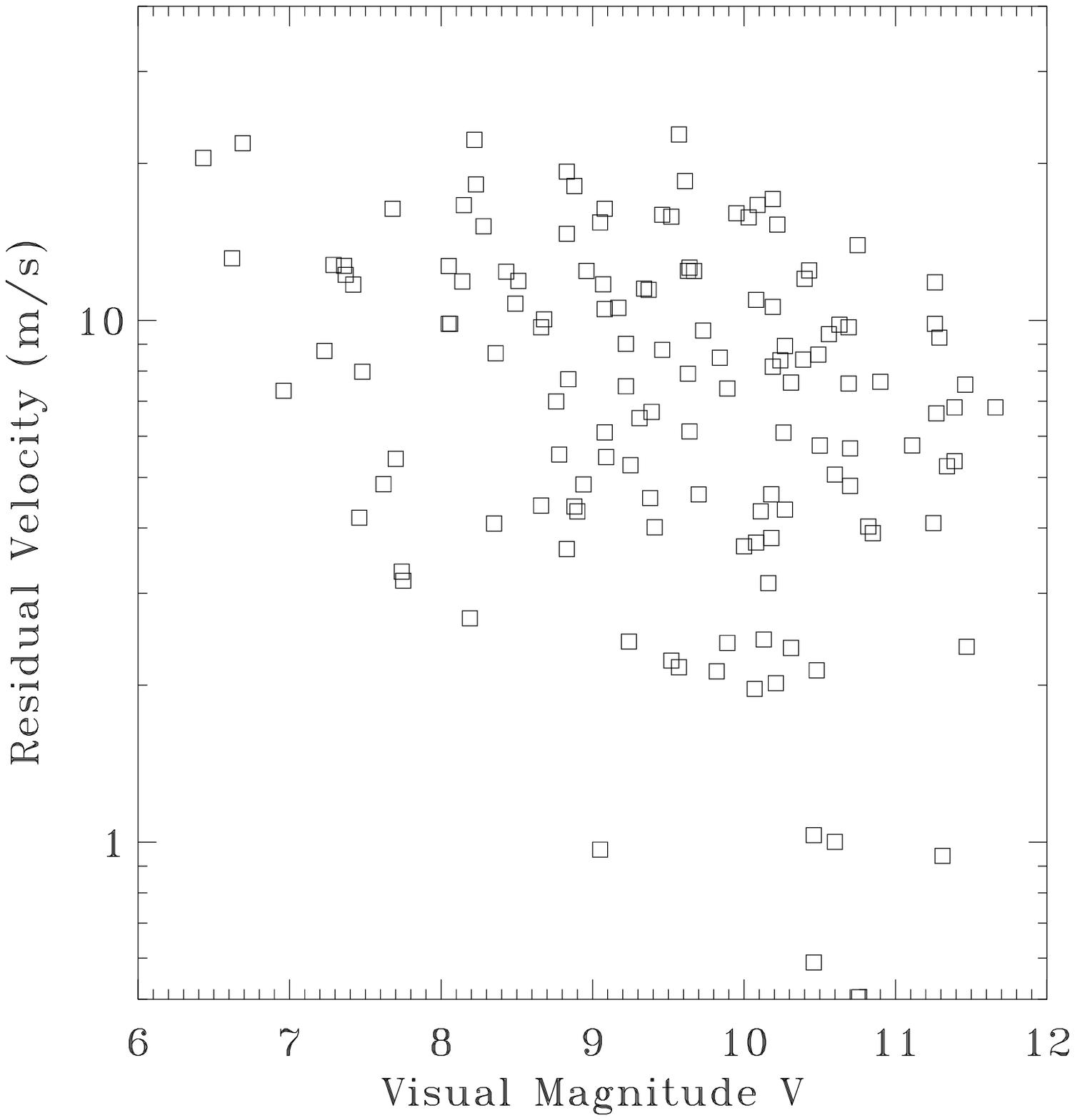} 
\end{array} $
\label{trends}
\caption{Radial velocity residuals (excluding variables) 
as a function of [Fe/H] (left), $T_\mathrm{eff}$ (center), and $V$ (right).}
\end{figure*}
Our analysis pipeline encompasses the full 
modeling of temporal and spatial variations of the instrumental profile 
of the HIRES spectrograph (Valenti et al. 1995) and is conceptually 
similar to that described by Butler et al. (1996). This analysis technique 
has allowed us to significantly improve upon our initial estimates of 
achievable radial velocity precision. 
In Figure 1, left panel, we show the histogram of the rms velocity 
residuals of the first 1.5 years of precise radial velocity measurements 
with HIRES for about 75\% of our sample. 
The rms velocitity residuals distribution 
of the {\it full} sample (excluding variables with rms $\ge 30$ m/s) 
averages $\sim 8$ m/s. For about two dozens of the stars in our sample, 
in common with the G dwarf planet survey of Latham (2000), we could 
establish the long-term stability of the velocity zero-point over 
time-scales of up to seven years (Figure 1, right panel). 
This demonstrates the true 
radial-velocity precision we are obtaining on the sample of metal-poor 
stars, with a significant improvement of $\sim 60$\% with respect 
to the targeted 20 m/s single-measurement precision. 


The exposure times predicted by the model derived from the simulations 
with the CfA library of stellar spectra are determined as a 
function of [Fe/H], $T_\mathrm{eff}$, and $V$. One possible matter 
of concern would be the evidence of systematic trends in the rms 
velocity distribution as a function of these three parameters. 
However, as shown in Figure 2, no clear rms velocity trends as a 
function of [Fe/H], $T_\mathrm{eff}$, and $V$ are present. This 
gives us confidence that the model we developed for the 
dependence of the radial velocity precision on the above 
parameters is robust. 
 

 \begin{figure}[!b]
 \begin{center}
    \leavevmode
 \centerline{\epsfig{file=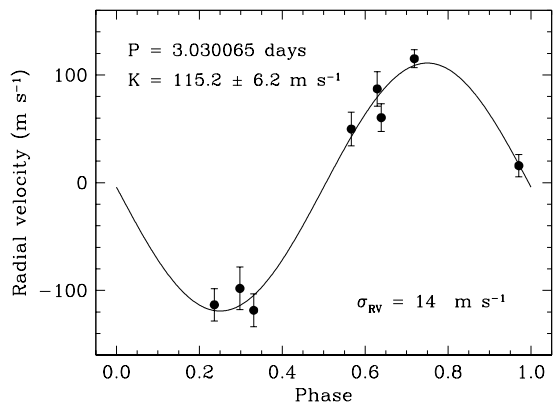,width=.9\linewidth}}
   \end{center}
  \caption{Radial velocity observations of TrES-1, 
overplotted with the best-fit orbit.}
  \label{rvtres1}
\end{figure}
We have provided an important confirmation of our ability 
to derive high-precision radial velocities for stars 2 to 5 mag 
fainter than the typical targets in Doppler surveys of nearby 
stars by determining the spectroscopic orbit for the recently 
announced (Alonso et al. 2004) transiting extra-solar planet TrES-1 
(Figure 3). As recently determined by means of detailed abudance 
analyses (Sozzetti et al. 2004), its parent star is a relatively cool 
($T_\mathrm{eff} \simeq 5250$ K), moderately faint (V = 11.79) 
solar-metallicity dwarf. The rms of the post-fit velocity residuals 
is $\sim 14$ m/s, in good agreement with the average of the internal 
errors. 
This is a remarkable result if we consider that this 
star is 2 mag fainter than the faintest stars with 
planets for which spectroscopic orbits with similar velocity 
precision have ever been derived.

 \begin{figure}[!b]
 \begin{center}
    \leavevmode
 \centerline{\epsfig{file=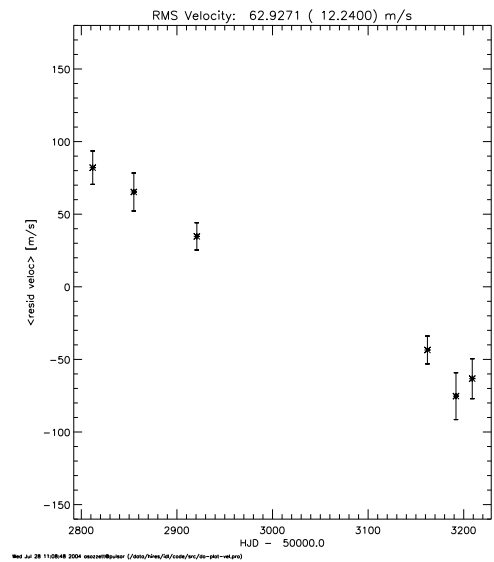,width=.8\linewidth}}
   \end{center}
  \caption{Radial velocity as a function of time for one of the 
variable objects in our sample.}
  \label{variab}
\end{figure}
\begin{figure*}[!t]
\centering
$\begin{array}{cc}
\includegraphics[width=0.40\textwidth]{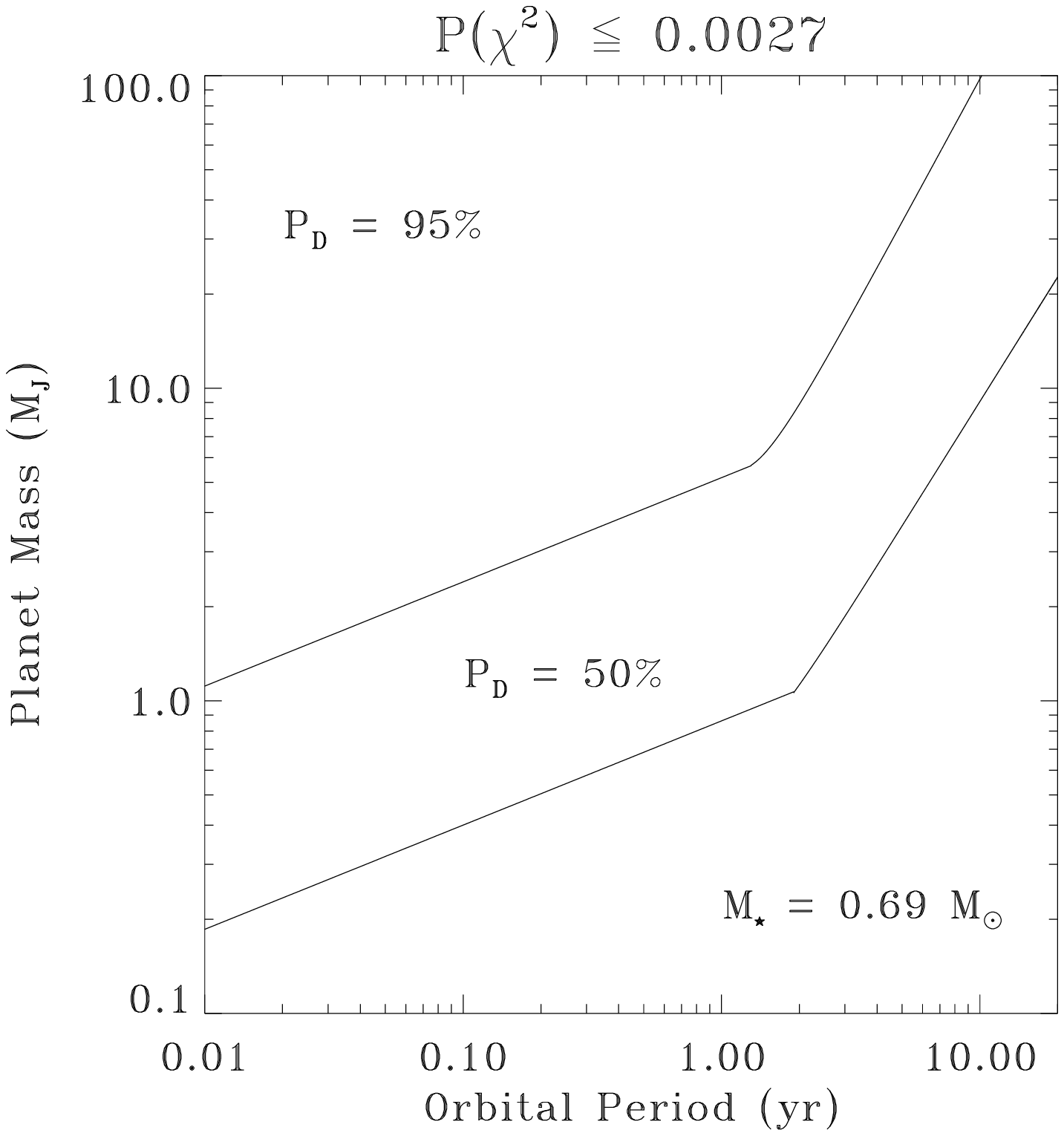} & 
\includegraphics[width=0.40\textwidth]{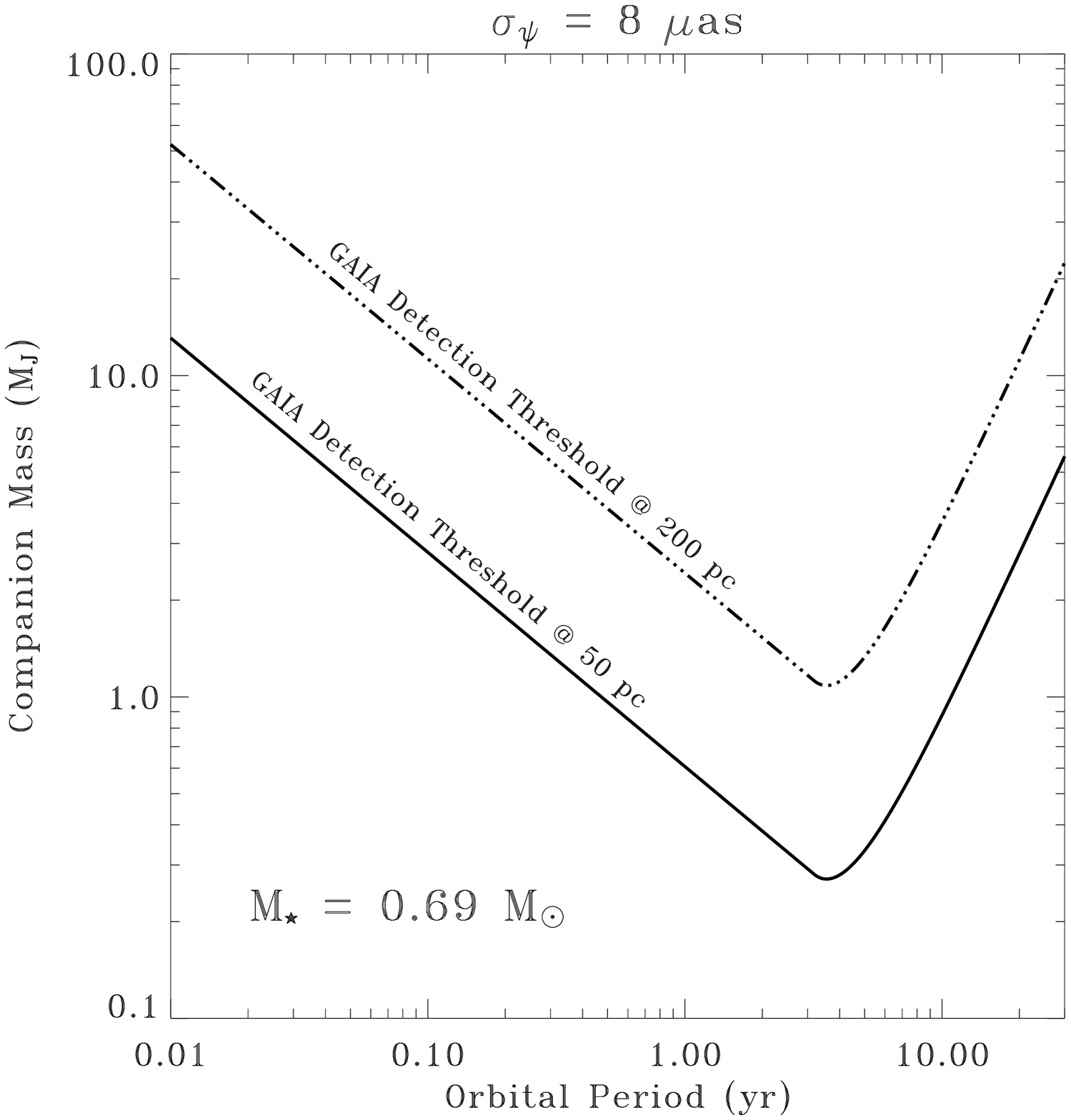} 
\end{array} $
\label{threshold}
\caption{Left: detectable planet mass as a function of orbital 
period for the sample of 149 metal-poor stars observed so far. 
Right: Gaia 95\% detection thresholds for bright ($V < 13$) 
metal-poor stars in the 50-200 pc distance range 
(based on the simulations of Sozzetti et al. 2003c).}
\end{figure*}

None of the 149 metal-poor dwarfs screened for planets so far 
(with an average number of 5 observations per target spanning 1.5 years), 
exhibits short-term, low-amplitude variations. However, $\sim 7\%$ of 
the stars in the sample appear to be long-period candidates. 
In Figure 4 we show radial velocities as a function of time 
for one of the objects with a large rms velocity value. 
The star shown in Figure 4 exhibits a linear radial-velocity trend 
which is indicative of the existence of a companion, and thus, 
together with another handful of objects, will become a primary target 
for follow-up observations. 


We have run Monte Carlo simulations to obtain a first estimate of 
the sensitivity of our survey to planetary companions of 
given mass and orbital period. In Figure 5, left panel, 
we show the minimum detectable planet mass $M_\mathrm{p}$ as a 
function of orbital period $P$, assuming a 0.69 $M_\odot$ primary 
(the average stellar mass of our sample) and 
a single-measurement precision $\sigma_\mathrm{RV} = 8$ m/s. 
For a typical observing strategy consisting of five observations 
spanning 1.5 years, the two curves identify the loci in the 
$M_\mathrm{p}-P$ discovery-space diagram for 50\% and 95\% 
probability of a 3-$\sigma$ detection, respectively. 
If present, essentially all Jupiter-sized objects on very-short 
periods ($< 1$ month) would have been detected. 
The data collected so far thus build toward evidence of the absence 
of close-in planets around metal-poor stars.  

The right panel of Figure 5 shows the $M_\mathrm{p}-P$ discovery 
space for the ESA Mission Gaia around a 0.69 $M_\odot$, bright 
($V < 13$) metal-poor star in the 50-200 pc distance range. 
A single-measurement error 
$\sigma_\psi = 8$ $\mu$as on the one-dimensional, along-scan 
coordinate and a 5-yr mission lifetime are assumed. 
The complementarity between Doppler and astrometric measurements 
is clearly evident. In particular, by surveying for planets all bright 
metal-poor dwarfs within 150-200 pc (of order of a few thousands), 
Gaia will help to understand 
whether the lack of close-in giant planets around the old stellar 
population extends also to the long-period regime, thus providing 
a firm statistical basis in favour of the core-accretion scenario for 
giant planet formation (e.g., Ida \& Lin 2004). Alternatively, the 
Gaia data, combined with ground-based radial-velocity monitoring, 
might confirm that metal-poor stars do harbor long-period giant planets 
(albeit at a reduced rate with respect to the metal-rich population), 
and thus different formation mechanims might have to be called 
into play (Boss 2002) and the role of metallicity on giant planet 
migration rates might have to be revisited (e.g., Sozzetti 2004). 

\section*{Acknowledgments}

A. S. gratefully acknowledges support from the Smithsonian Astrophysical 
Observatory through the SAO Predoctoral Fellowship program.

\end{document}